\documentclass{egpubl}
\usepackage{sca2026}
\usepackage{amsfonts}
\SpecialIssuePaper    

\arxiv

\usepackage[T1]{fontenc}
\usepackage{dfadobe}  

\usepackage{cite}  
\BibtexOrBiblatex
\electronicVersion
\PrintedOrElectronic
\ifpdf \usepackage[pdftex]{graphicx} \pdfcompresslevel=9
\else \usepackage[dvips]{graphicx} \fi

\usepackage{egweblnk} 

\usepackage{amsmath}
\usepackage{booktabs} 
\usepackage{wrapfig}
\usepackage{multirow}

\usepackage{mathtools}
\usepackage{bm}

\newcommand{\rev}[1]
 {
 \textcolor{black}{#1}
 }

\newcommand{\citet}[1]{\cite{#1}}
\let\oldparagraph\paragraph
\renewcommand{\paragraph}[1]{\oldparagraph*{#1}}

\title{Physics-Based Simulation of Contact-Induced Facial Wrinkling}

\author[Montes Maestre, et al.]
{\parbox{\textwidth}
    {\centering 
    J. S. Montes Maestre$^1$\orcid{0009-0001-9758-8861}, 
    L. Kavan$^2$\orcid{0000-0001-8549-0878},
    E. Boyer$^2$\orcid{0000-0002-1182-3729},
    R. Goldade$^2$\orcid{0009-0004-9336-8679},
    S. Coros$^1$\orcid{0000-0001-6604-4784}
    and B. Thomaszewski$^1$\orcid{0000-0001-8086-7664} 
    }
    \\
    {\parbox{\textwidth}
        {\centering 
        $^1$ETH Z\"urich, Switzerland\\
        $^2$Meta Reality Labs Research
       }
    }
}

\begin{document}

\teaser{
 \includegraphics[width=1.0\linewidth]{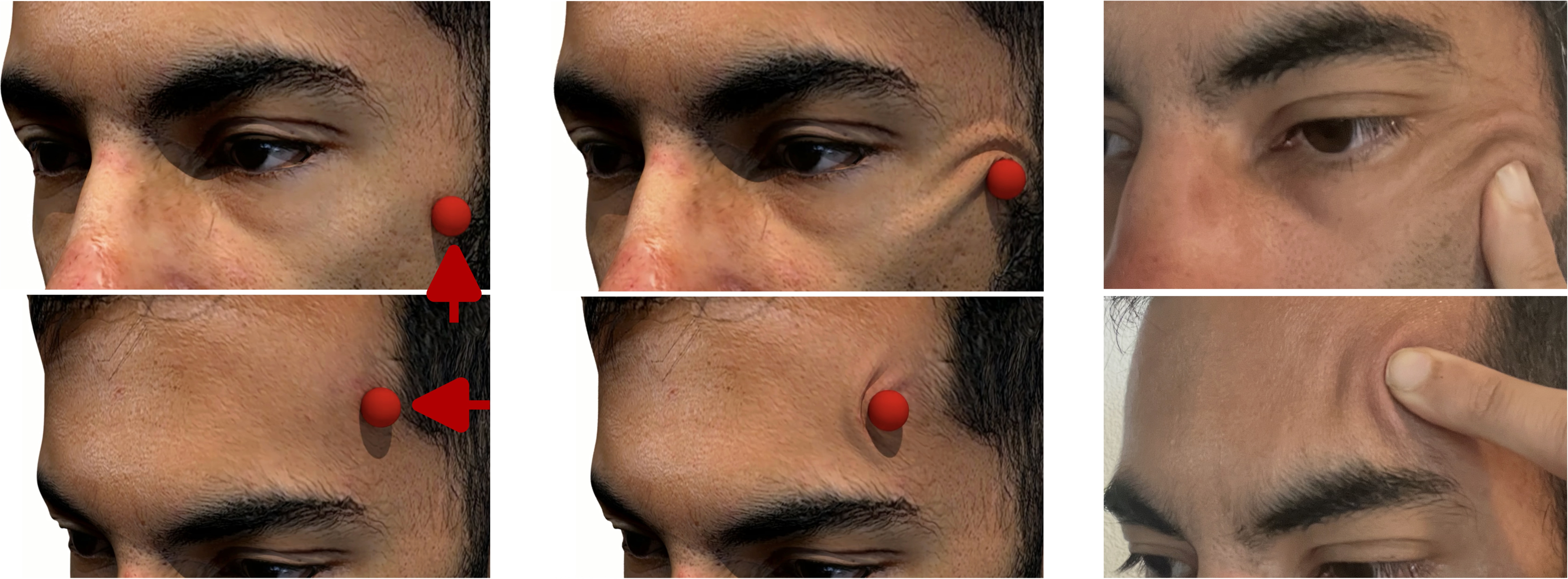}
 \centering
\caption{
\rev{Facial skin simulation on the temple (top) and forehead (bottom) regions. The initial states (left) are subjected to contact forces emulating the effect of a finger moving across the skin in the indicated direction (red arrows). Comparing against real-world reference footage (right), the simulated results demonstrate the model's ability to produce realistic wrinkling patterns with region-specific behaviors.
\vspace{1cm}
}
}
\label{fig:teaser}
}

\maketitle
\begin{abstract}
Facial skin dynamics are inherently challenging to simulate due to a combination of geometric, material, and anatomical complexities. Human skin is a nonlinear layered material with spatially heterogeneous attachments to the underlying tissues. During contact events, localized compression and shear induce mechanical instabilities, leading to fine-scale wrinkling patterns governed by a delicate interplay of geometry, boundary conditions, and through-the-thickness stresses. 

We present a finite element framework to simulate contact-induced wrinkling of facial skin. We model skin as a viscoelastic material with time-dependent relaxation that governs the rate, persistence, and damping of wrinkle formation. We employ high-order prismatic solid-shell elements to resolve through-thickness stresses and high-frequency deformation modes. Central to our approach, we introduce a continuum-based formulation of skin ligaments to model heterogeneous skin attachments and provide anatomically inspired mobility constraints. These skin ligaments control the formation and appearance of facial wrinkles by modulating their amplitude, wavelength, and spatial distribution. 

We evaluate our method on a set of synthetic examples and compare simulations with real‑world footage. These results demonstrate that our skin model produces temporally coherent and visually realistic wrinkle patterns during transient contact.


\end{abstract}  

\section{Introduction}

Realistic simulation of facial skin dynamics remains a challenge in computational biomechanics and computer graphics. Unlike generic soft bodies, facial skin is a structurally complex, multi-layered composite that exhibits highly non-linear behavior under deformation. Despite significant progress in simulating large-scale facial expressions, subtle transient mechanical interactions that occur during contact, such as a hand brushing against a cheek or a finger pressing into the forehead, remain difficult to capture. These interactions result in wrinkling phenomena that arise not only from the material properties of the tissue but also from the tissue's constrained mobility due to the skin's anatomical anchoring to underlying muscle and bone.

When skin is subjected to compression or shearing by external contact objects, local instabilities develop, manifesting as complex wrinkle patterns. These patterns are governed by a delicate balance between tissue stiffness, local geometry, and boundary conditions. Conventional thin-shell formulations often struggle to resolve these phenomena accurately because they neglect transverse shear and normal stresses through the thickness of the tissue. To capture the mechanics of contact-induced wrinkles, a simulation framework must resolve the stress state through-thickness of the skin while simultaneously respecting the heterogeneous mobility imposed by the fibrous ligaments connecting the dermis to the underlying fascia.

In this work, we develop a finite element model for contact-induced wrinkling of facial skin based on higher-order solid shell elements. Unlike standard shell elements, this discretization strategy captures the full volumetric structure of the skin, enabling distinct representations of the epidermis, dermis, and subcutaneous fat layers. Compared to conventional low-order elements, high-order basis functions enable the representation of high-frequency deformation modes, which are essential for the formation and propagation of fine-scale wrinkles during contact events.

As a central contribution of our work, we introduce \rev{a continuum-based formulation of skin ligaments} that model the anatomical attachment of skin to the underlying tissue. 
Facial skin is not uniformly attached to the skull: it slides over some regions while being tightly tethered in others. Explicit modeling of these attachments introduces localized constraints that govern the resulting wrinkle morphology, enabling us to achieve realistic and qualitatively accurate wrinkle formation.

Finally, the dynamic response of the tissue is essential for capturing the temporal evolution of wrinkles during sliding contact. We therefore model the skin using a viscoelastic formulation that captures the nonlinear and history-dependent behavior of biological tissues. Crucially, viscoelasticity enables temporal coherence, ensuring that wrinkle patterns do not emerge or vanish spuriously.

In summary, our proposed model provides an anatomically informed framework for studying how local contact, constrained mobility, and tissue viscoelasticity interact to generate complex wrinkle morphologies. We evaluate our method on a set of synthetic examples and perform ablation studies that validate the impact of various modeling choices and parameters. Comparisons between simulation results and real‑world footage demonstrate that our method is able to produce temporally coherent and visually realistic wrinkle patterns during sliding contact.

\section{Related work}

\paragraph{Digital Humans}
Creating life‑like digital representations of humans has long been a central goal of visual computing research. Over the past two decades, substantial progress has been made in the capture, reconstruction, and animation of personalized face \cite{egger20203d,blanz1999morphable,Beeler10HighQuality} and body models \cite{loper2015smpl,pons2015dyna,osman2020star}. More recently, digital humans have been augmented with increasing levels of biomechanical detail, such as bones \cite{Keller22Osso,keller2023skel}, organs \cite{Shetty23Boss,Guo22SMPLA}, and muscles \cite{komaritzan2021inside,AliHamadi13Anatomy}. These advances open the door to biomechanical simulation \cite{Zhang2023HACK,Lee09Comprehensive,Fan14Active} as a means of further enriching digital humans with increased realism, expressiveness, and predictive power. Our work contributes to this line of investigation with a novel anatomically-informed simulation model for facial wrinkling.

\textbf{Soft Tissue Simulation. }
Simulating the mechanics of soft tissue has been extensively studied in visual computing \cite{SoftTissueTeran2003,SoftTissueTeran2005,SoftTissueMcAdams2011,SoftTissueModi2020}. Within this domain, physics‑based simulation plays a central role in augmenting digital face models with realistic tissue deformations arising from facial expression, speech, and external contact \cite{Ichim2017,Kadlecek2019,Cong15Fully,Sifakis2005,Kozlov2017,Wagner2023,Yang2024}. Building on this body of work, we expand the capabilities of digital face models with a tailored approach for simulating contact‑induced wrinkling of facial skin.

\paragraph{Higher-Order Finite Elements. }

Despite their widespread use, low-order finite elements---and especially linear tetrahedra---are prone to artifacts such as locking and limited smoothness, particularly when applied to volume-preserving materials \cite{irving2007volume}. Numerous approaches have been proposed to alleviate these issues, including averaged volume constraints  \cite{irving2007volume,sheen2021volume}, mixed formulations \cite{francu2021locking,trusty2022mixed}, or higher-order finite elements \cite{Thomaszewski06AConsistent,Schneider22ALarge,Le23SecondOrder,Mezger08Interactive,martin2008polyhedral}. A recent line of work from the graphics community has explored prismatic solid-shell elements as an effective approach for modeling sheet materials that are thin but for which deformation in the thickness direction matters \cite{Chen23MultiLayer,Montes24,Montes23Differentiable}. As we demonstrate through our work, solid-shell elements are also a highly effective approach for modeling skin. In particular, we show that high-order basis functions are crucial for accurate capturing wrinkle formation.

\paragraph{Wrinkle Synthesis}
The mechanics of wrinkle formation have been extensively studied in the physics literature, where wrinkling is commonly understood as a mechanical instability arising from the interaction between material properties, geometry, and boundary conditions~\cite{genzer2006soft,cerda2003geometry}.
In the graphics community, wrinkles have often been addressed from a geometric or appearance-driven perspective. Prior work includes sketch-based approaches that enable interactive wrinkle synthesis~\cite{Kim15Interactive,rohmer2010animation,li2018foldsketch}, as well as procedural methods that augment coarse-scale simulations with fine-scale wrinkle detail~\cite{chen2023complex,muller2010wrinkle,Weiss23GraphBased,remillard2013embedded}.
Beyond these appearance-centric techniques, a body of work has focused on explicitly simulating the mechanics of skin wrinkling. Early approaches relied predominantly on mass-spring models~\cite{Koch96Simulating,Terzopoulos1990PhysicallybasedFM,MassSpringSkinWilhelms1997,MassSpringSkinAlbrecht2003,MassSpringSkinMurai2017}, while more recent efforts have transitioned toward finite-element formulations that offer improved physical fidelity~\cite{SkinTrianglesWu1996,Li:2013,magnenat2002computational}.
A related line of work investigates skin mechanics constrained to two-dimensional manifolds, such as surface-based models for facial or body skin~\cite{Li:2013}, as well as analogous formulations for cloth-like materials~\cite{Montes2020,Vechev2022}.
Our work is conceptually most closely related to the recent approach of Corigliano et~al.~\cite{corigliano2025viskin}, who model skin as an anisotropic elastic sheet augmented with ligament-like attachments. By prescribing the spatial distribution of ligament stiffness and rest length, their method achieves anatomically inspired, spatially varying skin mobility. While this approach produces realistic sliding behavior, it is inherently restricted to a two-dimensional representation and, unlike our method, cannot capture the volumetric mechanics required to predict skin wrinkling.

\section{Method}
\label{sec:method}

\begin{figure}[ht]
  \centering
   \includegraphics[width=1.\linewidth]{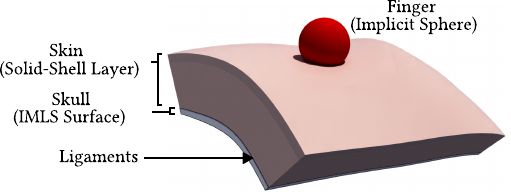}

   \caption{Computational model. The skin (epidermis and dermis) domain is discretized using a single layer of solid-shell elements. An implicit sphere represents the finger. The underlying rigid skull is modeled as an IMLS surface. A zero-length spring layer between the solid-shell skin and the IMLS surface represents the ligaments, governing the local attachment mechanics of the tissue relative to the skull.}
   \label{fig:skin_schematic}
\end{figure}

We develop a computational framework for simulating contact\-induced wrinkling of facial skin. We begin with a brief overview of skin anatomy (\S\ref{subsec:anatomy}), which motivates our finite‑element discretization based on high‑order prismatic solid‑shell elements (\S\ref{subsec:discretization}). The mechanical response of the skin is determined by a viscoelastic material model (\S\ref{subsec:material}). Contact between the skin and external objects is handled using $C^2$-continuous surface representations \rev{constructed via implicit moving least-squares (IMLS) or implicit functions} and penalty energies (\S\ref{subsec:contact}). To model the anatomical attachment of skin to the cranium, we introduce a novel continuum-based ligament formulation based on zero‑length Fung‑type springs (\S\ref{subsec:ligaments}). Finally, all components are integrated into a unified energy‑minimization framework (\S\ref{subsec:simulation}). \rev{\autoref{fig:skin_schematic} provides an overview of the full framework.}

\subsection{Anatomy of Human Skin}
\label{subsec:anatomy}
\begin{figure}
  \centering
   \includegraphics[width=1.\linewidth]{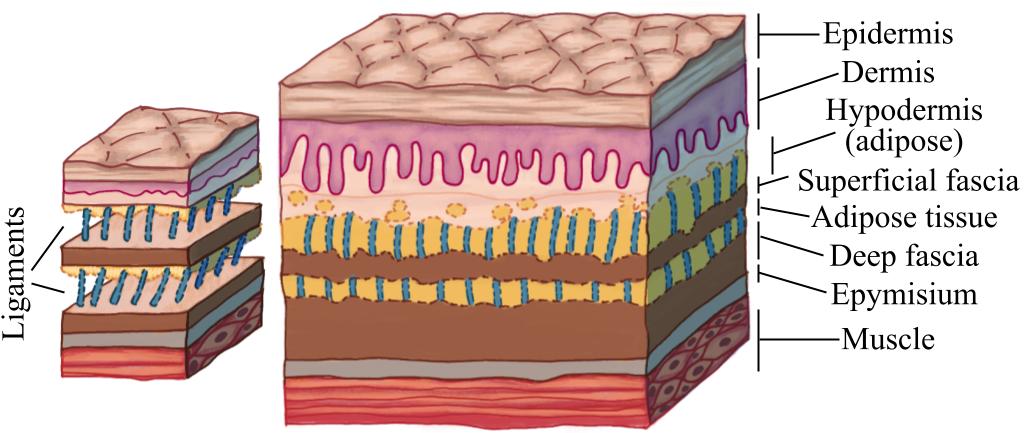}

   \caption{Skin structure (from \cite{corigliano2025viskin}).
   Right: skin classification in layers. Left: skin relations with neighboring structures. Skin ligaments crossing the adipose tissue are shown as blue threads.}
   \label{fig:skin_structure}
\end{figure}

Human skin is a complex organ composed of three primary layers: the epidermis, dermis, and hypodermis~\cite{Pissarenko2020}. As illustrated in \autoref{fig:skin_structure}, the \textit{epidermis} is the outer layer that serves as the interface between the body and the external environment. The \textit{dermis} contains dense networks of collagen and elastin fibers and plays a central role in determining the mechanical behavior of the skin. The \textit{hypodermis} is the inner layer that consists primarily of adipose tissue, which reduces friction with underlying structures and enables relative skin mobility. This adipose layer is traversed by \textit{skin ligaments}~\cite{Stecco2011,Nash2004} that anchor the skin to deeper tissues. These ligaments restrict arbitrary sliding, with the degree of mobility determined by their spatial density and length~\cite{Brenner2003}. In the following, we describe how this anatomical structure informs the design of our skin simulation model.


\subsection{Skin Model \& Discretization}
\label{subsec:discretization}


The anatomical complexity of human skin requires a carefully chosen level of abstraction when designing a computational model. 
A key observation is that the epidermis is significantly stiffer than the underlying tissues, and this difference in stiffness is a major driver of wrinkle formation \cite{genzer2006soft}. Therefore, we simplify the skin morphology into two layers: the epidermis and the underlying composite tissues that aggregate the dermis and hypodermis. 

\begin{figure}[b]
  \centering
   \includegraphics[width=1.\linewidth]{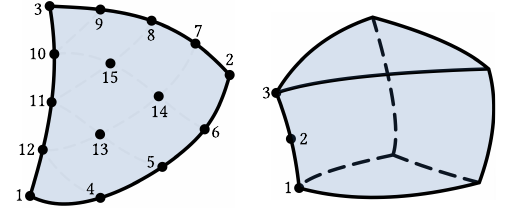}

   \caption{Solid-shell element configuration. The element is defined by a 15-node triangular base and a quadratic interpolation through the thickness, involving a total of 45 nodes.}
   \label{fig:shapeFunctions}
\end{figure}

To model this thin-layered but volumetric architecture of the skin, we employ prismatic solid-shell elements. Geometrically, these elements are defined by extruding a triangular base through the tissue thickness (see \autoref{fig:shapeFunctions}). We utilize a high-order interpolation scheme to prevent artificial numerical stiffening: quartic in-plane shape functions are used to resolve high-frequency wrinkles, while quadratic through-thickness interpolation is applied to mitigate volumetric locking, which is important for nearly incompressible biological materials.

\paragraph{Displacement Interpolation. }

We express the kinematics of prismatic solid shell elements in a natural coordinate system $\boldsymbol{\xi} = (u, v, w)$. The coordinates $u$ and $v$ parameterize the triangular cross-section, satisfying $u, v \ge 0$ and $u+v \le 1$, while $w \in [-1, 1]$ spans the through-the-thickness domain.  The position vector $\mathbf{x}$ at any point in the skin is computed via an isoparametric mapping,
\begin{equation}
    \mathbf{x}(u, v, w) = \sum_{I=1}^{n} N_I(u, v, w) \, \mathbf{x}_I \ ,
\end{equation}
where $\mathbf{x}_I$ are the nodal coordinates of the element. The shape functions $N_I:\mathbb{R}^3\rightarrow\mathbb{R}$ are constructed as the tensor product of the constituent bases,
\begin{equation}
N_{I(j,k)}(u, v, w) = P_j(u, v) \cdot T_k(w) \ ,
\end{equation}

\noindent where $P_j(u, v)$ denotes the set of $n_P=15$ quartic Lagrange polynomials defined over the triangular base and $T_k(w)$ represents the set of $n_T=3$ quadratic Lagrange polynomials along the thickness direction. Consequently, the element is composed of $n_P \times n_T = 45$ shape functions, where the global node index $I$ uniquely maps to a specific combination of in-plane index $j$ and through-thickness index $k$. This product structure ensures that the basis functions fully span the three-dimensional domain of the prism.

\paragraph{Deformation Gradient. }

The deformation of the skin is quantified by the deformation gradient $\mathbf{F}$, which maps changes in reference configuration $\mathbf{X}$ to \rev{changes} in the current configuration $\mathbf{x}$. Because shape functions are defined in the natural coordinate system $\boldsymbol{\xi}$, $\mathbf{F}$ is evaluated using the chain rule to relate spatial derivatives to the parametric domain,
\begin{equation}
\mathbf{F} = \frac{\partial \mathbf{x}}{\partial \mathbf{X}} = \frac{\partial \mathbf{x}}{\partial \boldsymbol{\xi}} \left( \frac{\partial \mathbf{X}}{\partial \boldsymbol{\xi}} \right)^{-1} \ .
\end{equation}

\paragraph{Volumetric-Isochoric Split. }

Given the nearly incompressible nature of biological tissue, it is computationally advantageous to decouple the deformation modes into \rev{shape-changing (isochoric) and volume-changing components}. To this end, we perform a multiplicative decomposition of the deformation gradient $\mathbf{F}$. Representing the local volume change as $J = \det(\mathbf{F})$, the isochoric component $\bar{\mathbf{F}}$ is defined as
\begin{equation}
\bar{\mathbf{F}} = J^{-1/3} \mathbf{F} \ .
\end{equation}
By construction, the isochoric component has unit determinant, $\det(\bar{\mathbf{F}}) = 1$. This decomposition ensures that $\bar{\mathbf{F}}$ only describes pure distortion, while $J$ exclusively describes volume change, allowing the strain energy density to be divided into their respective contributions.

\paragraph{Strain}

We measure deformation using the logarithmic Hencky strain tensor $\mathbf{E}$ and its isochoric counterpart $\bar{\mathbf{E}}$. This measure provides a natural extension of the linear strain to the large deformation regime, ensuring a symmetric and physically consistent response during extension and compression. Using the right Cauchy-Green tensor $\mathbf{C}=\mathbf{F}^{\text{T}} \mathbf{F}$,
the logarithmic strain tensor is defined as
\begin{equation}
\mathbf{E} = \frac{1}{2} \ln(\mathbf{C}) \ ,
\end{equation}
\rev{from which the isochoric strain follows  as 
\begin{equation}
   \bar{\mathbf{E}} = \mathbf{E} - \frac{1}{3}\ln(J)\mathbf{I} \ . 
\end{equation}
}
\subsection{Material Modeling}
\label{subsec:material}

\begin{figure}[b]
  \centering
   \includegraphics[width=1.\linewidth]{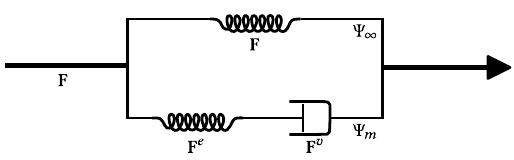}

   \caption{\rev{Spring-and-damper picture of the tissue model. A single spring ($\Psi_\infty$) sets the lasting stiffness, while a spring paired with a damper ($\Psi_m$) adds resistance that fades over time as the damper gradually takes up the deformation from its spring. Both terms share the same deformation $\mathbf{F}$.}}
   \label{fig:material-network}
\end{figure}

\rev{Soft facial tissue is viscoelastic: under sustained deformation its resistive stress gradually fades over time, and upon release it returns to its rest configuration slowly rather than recoiling instantaneously. To capture this behavior, we adopt a standard Generalized Maxwell model, which can be pictured as a spring and damper connected in series, set alongside a separate spring that is always engaged (see \autoref{fig:material-network}). While the deformation is held, the damper allows the spring to slowly relax and its restoring force gradually decreases. The separate spring keeps carrying the load. We express this conceptual model through its stored energy, written as the sum of three terms: a steady elastic term ($\Psi_{\infty}$) that sets the tissue's lasting stiffness, a dissipative term ($\Psi_\text{m}$) whose resistance fades over time, and a volumetric term ($\Psi_\text{vol}$) that penalizes changes in volume. Note that the first two terms, shown as the spring–damper network in \autoref{fig:material-network}, act exclusively on isochoric deformation, while the third enforces the near-incompressibility of the tissue.}

\rev{The total energy density of the skin $\Psi_{\text{skin}}$ is defined as the sum of these three terms,}
\rev{\begin{equation}
\Psi_{\text{skin}} = \Psi_{\infty} + \Psi_m + \Psi_{\text{vol}} \ .
\label{eq: skin energy}
\end{equation}}
\rev{We define each term below.}


\paragraph{Steady Elastic Energy Term}

\rev{The steady elastic term represents the long-term, non-relaxing stiffness of the tissue}. We model this using a Fung-type strain energy density $\Psi_{\infty}$ based on the isochoric logarithmic strain $\bar{\mathbf{E}}$. This formulation accounts for the rapid ``strain-stiffening'' behavior typical of biological membranes \cite{fung1981biomechanics},
\begin{equation}
\Psi_{\infty}(\bar{\mathbf{E}}) = \frac{\mu_\infty}{C} \left( \exp \left[ C \text{tr}(\bar{\mathbf{E}}^2) \right] - 1 \right) \ ,
\end{equation}
where $\mu_\infty$ and $C$ are material constants that govern the initial stiffness (linear regime) and the exponential stiffening rate, respectively.

\paragraph{Dissipative Energy Term}
\label{sec:dissipative}

\rev{To model the relaxation of the tissue, we split the deformation gradient multiplicatively into an elastic component $\mathbf{F}^e$ (spring) and a viscous component $\mathbf{F}^v$ (damper),}
\begin{equation}
\mathbf{F} = \mathbf{F}^e \mathbf{F}^v \text{,}
\end{equation}
\rev{where $\mathbf{F}$ is the standard deformation gradient, obtained directly from the displacement field, and $\mathbf{F}^v$ is an internal state variable tracking the viscous history. The deformation gradient of the elastic component is then recovered as $\mathbf{F}^e = \mathbf{F}\ (\mathbf{F}^v)^{-1}$. }

\rev{To describe the energy stored within the elastic spring element, we implement an isochoric Saint Venant–Kirchhoff (StVK) energy density potential. Formulated in terms of the isochoric elastic logarithmic strain $\bar{\mathbf{E}}^e$, the potential is expressed as}
\begin{equation} 
\Psi_m(\bar{\mathbf{E}}^e) = \mu_m \operatorname{tr}\left([\bar{\mathbf{E}}^e]^2\right) \ ,
\end{equation}
\rev{where $\mu_m$ represents the elastic modulus of the spring. Because the isochoric logarithmic strain is  traceless ($\text{tr}(\bar{\mathbf{E}}^e) = 0$), any corresponding volumetric energy term vanishes. At the onset of loading, the damper has not yet relaxed ($\mathbf{F}^v = \mathbf{I}$), meaning that the elastic component temporarily carries the entirety of the deformation ($\mathbf{F}^e = \mathbf{F}$). Over time, however, the damper gradually absorbs the isochoric portion of the deformation ($\mathbf{F}^v \to \bar{\mathbf{F}}$), which subsequently reduces $\mathbf{F}^e$ strictly to pure volumetric deformation. Because the energy potential $\Psi_m$ depends solely on this isochoric elastic strain, the system naturally drives toward a zero-energy state as the elastic deformation becomes purely volumetric, ultimately leading to the limits $\bar{\mathbf{E}}^e \to \mathbf{0}$ and $\Psi_m \to 0$. Consequently, the dissipative term eventually ceases to contribute to the macroscopic stress response. The exact kinetic law governing the evolution of $\mathbf{F}^v$ over time is detailed in Section \ref{subsec:simulation}.}

\paragraph{Volumetric Penalty}

To account for the nearly incompressible nature of the tissue, we introduce a volumetric penalty term,
\begin{equation}
\Psi_{\text{vol}}(J) = \frac{1}{2} K (J - 1)^2 \ ,
\end{equation}
where $K$ denotes the bulk modulus.

\paragraph{Multilayered Material Integration}

\rev{
To capture the stratified architecture of the skin—thin, stiff epidermis and soft underlying tissue (dermis and fat)—a separate layer of elements could be used for each anatomical layer. However, this approach would lead to unmanageably large problem sizes and long computation times. 
We instead use a single layer of elements combined with a composite integration scheme that accounts for material variation through the thickness. To this end, we partition the through-the-thickness domain into two distinct sub-domains corresponding to the stiff epidermis and the soft tissue layer below. To  capture the sharp mechanical contrast necessary for surface wrinkling, we employ a full Gaussian quadrature rule for each sub-domain.
}
The total strain energy $U_{\text{skin}}$ is calculated as the summation of the skin strain energies, defined in (\ref{eq: skin energy}), contributed by each skin layer $l$,
\begin{equation}
U_{\text{skin}} = \sum_{l \in {\text{epi, soft}}} 
\int_{\Omega_l} \Psi_{\text{skin}, l}\ d\Omega \ .
\end{equation} 

\subsection{Contact}
\label{subsec:contact}

Skin interactions are modeled using a penalty-based contact formulation, where external objects and the skull are represented as rigid implicit surfaces. This approach avoids irregularities in the tangential forces that arise in $C^0$ meshes~\cite{Du2024}. The contact condition is enforced through a gap function $g$, which represents the signed distance between the two bodies. To prevent non-physical interpenetration without adhesion artifacts, we define a one-sided penalty function
\begin{equation}
\Psi_{\text{contact}} = \begin{cases} \kappa g^2 & \text{if } g \le 0 \\ 0 & \text{if } g > 0 \end{cases} \ ,
\end{equation}
where $\kappa$ is a penalty stiffness parameter.

\paragraph{Fingertips} External collision objects, such as a fingertip, are modeled using an implicit sphere, 
\begin{equation}
\label{eq:sphere-collider}
g_{\text{sphere}}(\mathbf{x}) = \|\mathbf{x} - \mathbf{x}_c\| - R \ ,
\end{equation}
where $\mathbf{x}_c$ is the center and $R$ is the radius of the sphere. This ensures a smooth localized pressure distribution in the epidermis.

\paragraph{Skull} We use a smooth surface representation to ensure that the soft tissue layer can slide freely over the underlying bone structure.
We represent the skull geometry through a point cloud $\mathcal{P} = \{ \mathbf{p}_i \}$ and a corresponding set of unit normals $\mathbf{n}_i$, extracted directly from the interior surface of the initial face mesh. Following the Implicit Moving Least Squares (IMLS) approach, the skull surface is represented as the zero-level set of a smooth, $C^2$-continuous implicit function, which serves as the gap function $g_{\text{skull}}$ for the internal contact interface,
\begin{equation}
g_{\text{skull}}(\mathbf{x}) = \frac{\sum_{i \in \Omega(\mathbf{x})} \theta(\|\mathbf{x} - \mathbf{p}_i\|) \langle \mathbf{n}_i, \mathbf{x} - \mathbf{p}_i \rangle}{\sum_{i \in \Omega(\mathbf{x})} \theta(\|\mathbf{x} - \mathbf{p}_i\|)} \ ,
\end{equation}
\noindent where 
\begin{equation}
\theta(d) = \begin{cases} (1 - \frac{d}{h})^4 & \text{if } d < h \\ 0 & \text{if } d \ge h \end{cases}
\label{eq:localfunc}
\end{equation}
\noindent is a compactly supported radial weight function and $\Omega(\mathbf{x})$ denotes the local neighborhood of points where $\theta(d) > 0$. 

\paragraph{Friction}

We integrate friction into our variational framework through a dissipative potential density $\Psi_\text{friction}(\delta)$, where $\delta = \|\mathbf{v}(\mathbf{x})\|$ is the relative tangential velocity magnitude. Following Li et al. \cite{Li20IPC}, we use the lagged normal force $f_n$ from the previous converged timestep. The corresponding potential density is defined as
\begin{equation}
\Psi_{\text{friction}}(\delta) = \mu f_n \cdot \begin{cases} \frac{\delta^6}{\epsilon^5} - \frac{3\delta^5}{\epsilon^4} + \frac{5\delta^4}{2\epsilon^3} & \text{if } \delta < \epsilon \\ \delta - \frac{\epsilon}{2} & \text{if } \delta \ge \epsilon \end{cases} \text{.}
\end{equation}
\noindent In the sticking regime ($\delta < \epsilon$), the high-order polynomial provides a smooth, zero-gradient start at $\delta=0$. In the slipping regime ($\delta \ge \epsilon$), the potential becomes linear, resulting in a constant force magnitude $\mu f_n$. This specific formulation ensures $C^2$-continuity across the transition, eliminating numerical singularities and providing a stable foundation for the solver.

\paragraph{Numerical Integration}

Standard quadrature rules often fail to adequately model the pressure distribution when there is a significant geometric mismatch between the skin mesh and the implicit colliders (the sphere and the IMLS skull). To address this problem, we increase the density of integration points on the contact surfaces, effectively performing a more fine-grained sampling of the gap function $g$ and the frictional potential $U_{\text{friction}}$. This refinement prevents aliasing artifacts in which the skin might locally penetrate an implicit surface between sparsely distributed nodes.

\subsection{Ligaments}
\label{subsec:ligaments}

The mechanical anchoring of skin tissue is governed by a Fung-type strain energy potential, $\Psi_{\text{ligament}}$, integrated over the area of the inner shell surface. Because this surface is initialized in direct contact with the implicit skull surface, we model ligaments as zero-length springs where the displacement $\delta = \|\mathbf{x} - \mathbf{X_\text{skull}}\|$ characterizes the deviation from the rest state. The potential is defined as
\begin{equation}
\Psi_{\text{ligaments}}(\delta) = \frac{k}{b(\mathbf{x})} \left( e^{b(\mathbf{x})\delta} - 1 \right) \ ,
\end{equation}
where $k$ is the base stiffness scaling factor and $b(\mathbf{x})$ is a dimensionless parameter determined by a spatially varying heat map. By modulating $b(\mathbf{x})$, we can precisely control anatomical "pinning" points, using high values to restrict skin mobility and lower values to allow for sliding of the soft tissue over the skull.

\paragraph{Ligament Heat Map}


We construct the continuous ligament stiffness field from discrete, empirically determined source locations using a variational diffusion framework based on the Vector Heat Method \cite{Sharp2019}. We define two auxiliary scalar fields: an unnormalized stiffness field $\hat{b}_t$, and a magnitude-normalization field $\phi_t$, where $t$ is the diffusion time. At each discrete source point location, $\hat{b}_0$ is initialized with the desired ligament stiffness, while the normalization field is set to $\phi_0=1$. To ensure compatibility with high-order shape functions and prevent numerical artifacts (e.g., negative gradients on high-order nodes), we define the initial fields as a weighted sum of sources using the locally supported smoothing function (\ref{eq:localfunc}). The time-evolved auxiliary fields $\hat{b}_t$ and $\phi_t$ are computed by solving heat equations, and the ligament stiffness heatmap is then computed as the node-wise quotient of the auxiliary fields,
\begin{equation}
    b = \frac{\hat{b}_t}{\phi_t} \ .
\end{equation}

\subsection{Time Stepping}
\label{subsec:simulation}

\rev{The system is advanced in time using an implicit Euler scheme. For a time increment $\Delta t$, the updated configuration $\mathbf{x}_{i+1}$ and internal viscous variables $\mathbf{F}^v_{i+1}$ are determined by minimizing the total incremental potential~\cite{Gast2015,Martin2011},
\begin{equation}
\label{eq:incremental-potential}
\begin{aligned}
\mathbf{x}_{i+1}, \mathbf{F}^v_{i+1} = \arg\min_{\mathbf{x}, \mathbf{F}^v} \Big( &U_{\text{skin}}(\mathbf{x}, \mathbf{F}^v) + U_{\text{contact}}(\mathbf{x}) \\
&+\, U_{\text{friction}}(\mathbf{x}) + U_{\text{ligaments}}(\mathbf{x})\Big)\ ,
\end{aligned}
\end{equation}
where $U_{\text{contact}}$ and $U_{\text{friction}}$ are updated at each step by translating the centers $\mathbf{x}_c$ of the implicit collision spheres to track the prescribed kinematics of the fingers. While the system exhibits rate-dependent dynamic behavior via its viscous components, inertial effects are assumed to be negligible and are omitted from the formulation. The discrete unknowns to resolve are the spatial positions $\mathbf{x}$ and the local viscous variables $\mathbf{F}^v$ at each quadrature point. Because $\mathbf{F}^v$ enters the energy exclusively through the dissipative term $\Psi_m$, its temporal evolution can be decoupled from the global spatial degrees of freedom and resolved via a constitutive flow law. }

\paragraph{Viscous Time Integration}

\rev{The damper follows a linear relaxation law: its viscous strain grows at a rate
proportional to the strain currently carried by the spring,
\begin{equation}
  \dot{\bar{\mathbf{E}}}^v = \frac{1}{\tau}\,\bar{\mathbf{E}}^e \ ,
  \label{eq:relax-ode}
\end{equation}
where $\tau$ is the relaxation rate of the material. To advance from step $i$ to step $i+1$, we hold the damper fixed at its previous
state $\mathbf{F}^v_i$ and form the \emph{trial elastic gradient}, the stretch the
spring would carry if the damper did not move during the step,
\begin{equation}
  \bar{\mathbf{F}}^{e,\mathrm{tr}} = \bar{\mathbf{F}}_{i+1}\,(\mathbf{F}^v_i)^{-1},
  \label{eq:trial-gradient}
\end{equation}
whose principal stretches and directions we extract via SVD,
$\bar{\mathbf{F}}^{e,\mathrm{tr}} = \mathbf{U}_e\bar{\boldsymbol{\Sigma}}_e\mathbf{V}_e^T$ \ .}

\rev{We assume that the viscous increment (the deformation the damper takes up during the
step) is coaxial with this trial elastic state, sharing the same principal axes
$\mathbf{V}_e$ (an assumption on the increment alone, not on the full deformation history). Along each principal axis, the one-dimensional analogy applies: multiplying
stretches is equivalent to adding log-strains, just as $\ln(ab) = \ln(a)+\ln(b)$.
Since the damper absorbs from the spring at a rate of $\tfrac{1}{\tau}\bar{\mathbf{E}}^e$,
holding the total deformation fixed at its trial value gives
$\dot{\bar{\mathbf{E}}}^e = -\tfrac{1}{\tau}\bar{\mathbf{E}}^e$,
whose exact solution is
$\bar{\mathbf{E}}^e(t) = \bar{\mathbf{E}}^{e,\mathrm{tr}}\,e^{-t/\tau}$.
After a step $\Delta t$, defining $\alpha = e^{-\Delta t/\tau}$, the spring's
log-strain decays by $\alpha$ along each principal axis, so the damper absorbs the
complementary fraction,
\begin{equation}
  \Delta\bar{\mathbf{E}}^v = (1-\alpha)\,\bar{\mathbf{E}}^{e,\mathrm{tr}},
\end{equation}
and the remaining fraction is elastic after the step,
\begin{equation}
  \bar{\mathbf{E}}^e_{i+1} = \alpha\,\bar{\mathbf{E}}^{e,\mathrm{tr}} .
\end{equation}}

\rev{The spring now holds the strain $\alpha\,\bar{\mathbf{E}}^{e,\mathrm{tr}}$, so the
stress it exerts is
\begin{equation}
  \frac{\partial \Psi_m}{\partial \bar{\mathbf{E}}^e}
  \bigg|_{\alpha\,\bar{\mathbf{E}}^{e,\mathrm{tr}}}
  \;=\; 2\mu_m\,\alpha\,\bar{\mathbf{E}}^{e,\mathrm{tr}}
  \;=\; \alpha\,\frac{\partial \Psi_m(\bar{\mathbf{E}}^{e,\mathrm{tr}})}
                     {\partial \bar{\mathbf{E}}^{e,\mathrm{tr}}}.
\end{equation}
This is the gradient of $\alpha\,\Psi_m(\bar{\mathbf{E}}^{e,\mathrm{tr}})$ with
respect to the trial elastic strain, so the incremental energy density for the dissipative
term is
\begin{equation}
  \Psi_m^{\mathrm{relax}} = \alpha\,\Psi_m\!\left(\bar{\mathbf{E}}^{e,\mathrm{tr}}\right).
  \label{eq:maxwell-relaxed}
\end{equation}
This energy density contributes alongside $\Psi_\infty$ and $\Psi_{\mathrm{vol}}$ to the
total $\Psi_\mathrm{skin}$ in Eq.~\eqref{eq:incremental-potential}; once the minimizer $\mathbf{x}_{i+1}$
is found, the damper is advanced by applying the absorbed fraction $1-\alpha$ to the
principal stretches of the trial elastic state,
\begin{equation}
  \mathbf{F}^v_{i+1} = \Bigl( \mathbf{V}_e\,
  \bar{\boldsymbol{\Sigma}}_e^{\,1-\alpha}\,
  \mathbf{V}_e^{T} \Bigr)\,\mathbf{F}^v_i \ .
  \label{eq:viscous-update}
\end{equation}}

\paragraph{Non-Linear Solver} 
\rev{For high-order elements, computing and factorizing the system matrix (i.e., the energy Hessian) is the computationally most intense part of a Newton step when solving for the updated positions $\mathbf{x}_{i+1}$. For this reason, we compute and factorize the system's Hessian once at the beginning of the simulation and subsequently apply L-BFGS updates to approximate the inverse Hessian. To ensure robust convergence for this quasi-Newton method, we recompute and refactorize the Hessian every 50 iterations or if the step size returned by line search falls below $10^{-6}$. We consider the system converged, once the gradient norm falls below $10^{-6}$.}
\section{Results}
\label{sec:results}

In this section, we evaluate the performance of our framework by analyzing how individual mechanical parameters influence the emergent shape of skin wrinkles. We first explore the sensitivity of the model to ligament stiffness, observing how the internal tethering of the tissue induces changes in the amplitude and frequency of the wrinkles. We then proceed to analyze the interlayer stiffness ratio between the epidermis and the underlying tissue, which we find to be an important contributor to high-frequency surface wrinkles. To capture the time-dependent nature of skin deformation, we examine the effect of viscoelasticity, specifically focusing on how stress relaxation influences the dynamic behavior of wrinkles under contact. Finally, we provide a qualitative comparison between our simulated results and real-world footage, demonstrating that the interplay of these mechanical factors enables the model to replicate the complex, dynamic wrinkling patterns observed in real faces.

\rev{We follow ~\cite{Feng2022} and set Young's modulus
to $4\,\text{MPa}$ for the epidermis and $20\,\text{kPa}$ for the underlying tissue,
giving a stiffness ratio of $200:1$. Both layers share a Poisson's ratio of
$\nu = 0.49$, chosen to enforce near-incompressibility. The stiffening rate of the
steady elastic term is set to $C = 1$ for both layers. The shear modulus
$\mu$ and bulk modulus $K$ are then derived from standard
linear elastic relations,
\begin{equation}
  \mu = \frac{E}{2(1+\nu)},
  \qquad
  K = \frac{E}{3(1-2\nu)} \ .
  \label{eq:material-params}
\end{equation} 
$\mu$ is then split between the steady and dissipative terms $\mu_\infty=0.2\mu$ and $\mu_m=0.8\mu$. The time-dependent response is defined by a viscoelastic relaxation time $\tau = \frac{1}{12}s$ and a simulation step size of $\Delta t = \frac{1}{24}s$. For contact, we use a penalty parameter of $\kappa=10$ MPa and a friction coefficient of $\mu = 0.8$.}

\paragraph{Quadrature Rules}

\rev{Numerical integration uses the fully symmetric, positive-weight rules of~\cite{Witherden2015}: the 85-point degree-10 prism rule for skin layers, to densely sample the spatially varying viscous deformation gradient, and the 25-point degree-10 triangle rule for contact integrals. For the elastic sheet experiments in \autoref{fig:wrinkled-sheet}, sparser rules suffice: 11 points at degree 4 for quadratic elements, 16 points at degree 5 for cubic, and 29 points at degree 6 for quartic. Linear elements use the 2-point rule of~\cite{Chen23MultiLayer}. Precise rule data are provided in the supplementary material of~\cite{Witherden2015}.}

\subsection{Ligament Stiffness}

\begin{figure}[t]
  \centering
   \includegraphics[width=1.\linewidth]{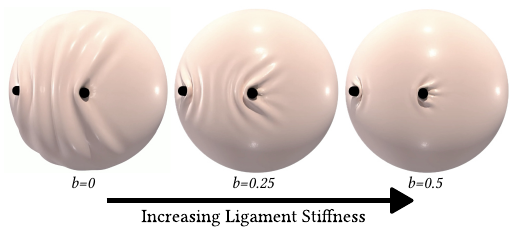}
   \caption{Impact of ligament stiffness parameter $b$ on wrinkle morphology. 
   As ligament stiffness increases (left to right), the mobility of the skin is progressively restricted.}
   \label{fig:ligament-stiffness}
\end{figure}

The ligament stiffness acts as a fundamental control parameter for the spatial frequency and amplitude of the skin's buckling patterns. \autoref{fig:ligament-stiffness} highlights the impact of ligament stiffness on wrinkle morphology. Low-stiffness ligaments result in skin that is loosely tethered, allowing compressive stresses to spread over a wide domain and resulting in large-scale, long-wavelength wrinkles that span extensive areas of the surface. As stiffness increases, ligaments progressively restrict the movement of the skin to smaller, constrained regions. This constraint on movement forces the skin into a higher energy state characterized by higher frequency, lower amplitude wrinkles that are tightly localized around the contact points. Finally, as the stiffness reaches the limit of full attachment, the skin is effectively pinned to the substrate, causing wrinkles to largely disappear.


\subsection{Interlayer Stiffness Ratio}

\begin{figure}[t]
  \centering
   \includegraphics[width=1.\linewidth]{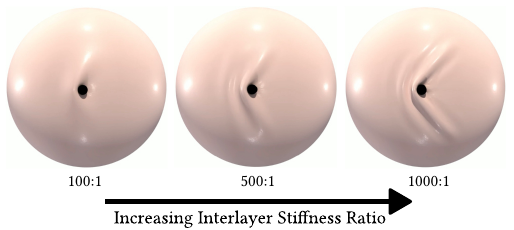}
   \caption{Sensitivity to interlayer stiffness. Increasing the stiffness ratio between skin layers results in finer wrinkling details.}
   \label{fig:ligament-ratio}
\end{figure}

Similar to ligament stiffness, the stiffness ratio between skin layers (i.e., the stiff epidermis and the more compliant underlying substrate) also influences wrinkle frequency. \autoref{fig:ligament-ratio} demonstrates that a low ratio ($100:1$) produces a smooth response where the layers deform in a unified bulk manner. In contrast, a larger stiffness ratio $(1000:1)$ amplifies the disparity between epidermis and substrate, leading to surface instability and the formation of high-frequency wrinkles. These observations suggest that ligament stiffness controls the macro-structure of the folds whereas the interlayer ratio modulates the frequency.

\subsection{Viscoelasticity}

Viscoelasticity influences how human skin deforms under lateral pressure, a phenomenon we test in the {\textit{supplementary video}}. In the first simulation, we treat the skin as a fully elastic material, which results in a spring effect: the skin recovers instantly and tension is redistributed across the surface, causing the deformation to move smoothly with the contact point. In the second simulation, the viscoelastic model introduces a rate-dependent delay where internal friction prevents the skin from redistributing tension immediately. This delay creates a memory effect in the tissue, as seen in the real-world footage. Because the skin cannot relax instantly, once a wrinkle is formed, it is sustained in the tissue as the finger moves forward. This leads to a progressive accumulation of folds that are held in place by the material's internal resistance.

\subsection{Qualitative Evaluation}

\begin{figure}[b]
  \centering
   \includegraphics[width=1.\linewidth]{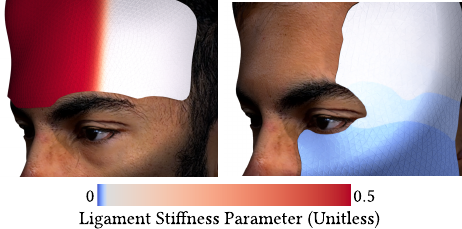}

   \caption{Spatial distribution of the ligament stiffness parameter $b$. The heatmaps illustrate the mapping of the exponential stiffness parameter $b$ across the forehead (left) and temple/cheek regions (right). Higher values (red) denote dense ligamentous anchoring. Moderate values (white) indicate transitional zones with balanced skin tethering. Lower values (blue) indicate areas of high skin mobility.}
   \label{fig:distribution}
\end{figure}

To evaluate the qualitative accuracy of our model, we collected several videos of a human subject creating facial wrinkling using their finger. We compare specific anatomical regions of the face (\autoref{fig:distribution}), such as the temples or the forehead, for localized simulation and evaluation. This setup allows us to maintain a significantly higher mesh resolution within the contact zone, ensuring that the buckling instabilities and dissipative effects are captured with the precision required to match the dynamics observed in the captured footage.


\begin{figure*}[h]
  \centering
   \includegraphics[width=1.\linewidth]{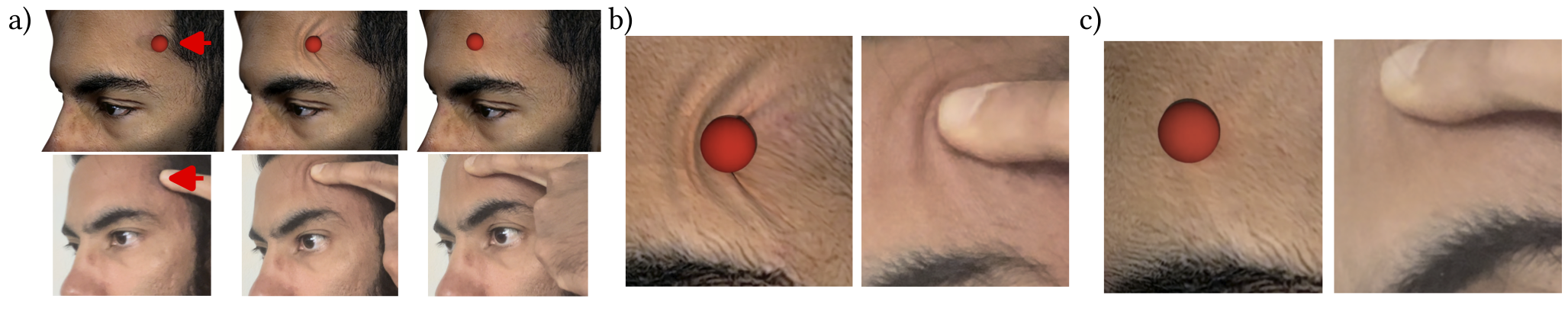}

   \caption{Comparison between forehead simulation and real-world reference. (a) Lateral displacement from the temple toward the center of the forehead; the simulation (top) captures the localized buckling and wrinkle accumulation seen in the video reference (bottom). (b) and (c) Medial displacement from the temple toward the center of the forehead. In (b), high-frequency wrinkles emerge in the lateral region where ligaments are moderately attached, whereas in (c), these wrinkles vanish as the skin reaches the strongly attached central region.}
   \label{fig:temporal}
\end{figure*}

First, we evaluate the temple and upper cheek area by simulating a finger dragging the skin upward (\autoref{fig:teaser}-{\textit{top}}). This region is characterized by a distinct gradient in ligament density: the cheek tissue is loose and lacks deep anchors, whereas the temple skin is relatively attached to the underlying tissues. As the finger drags the tissue, the loose cheek skin is easily displaced upward. As this displaced skin is forced toward the temple, it encounters the denser ligamentous anchors and begins to gather. This accumulation of tissue triggers a buckling response, resulting in the formation of a high-amplitude wrinkle pattern, which is captured by our model.

In our second experiment, we drag the skin horizontally from the temple toward the center of the forehead (\autoref{fig:teaser}-{\textit{bottom}}). In this region, the skin near the temples is relatively sparse in ligamentous attachments compared to the center of the forehead, where the tissue is much more firmly anchored to the underlying frontal bone. As the finger moves through the sparser temple-side area, the skin is easily displaced, allowing high-frequency wrinkles to form ahead of the contact point (\autoref{fig:temporal}-{\textit{b}}). However, as the finger progresses toward the center, it enters the zone with higher ligament density. At this stage, the skin is pulled taut against its central anchors, causing the wrinkles to vanish as the tissue reaches a state of high tension (\autoref{fig:temporal}-{\textit{c}}). This transition illustrates the model's ability to balance external contact forces with localized anatomical constraints to produce realistic, non-uniform skin behavior.

\subsection{Performance of High-Order Solid Shells}

\begin{figure*}[h]
  \centering
   \includegraphics[width=1.\linewidth]{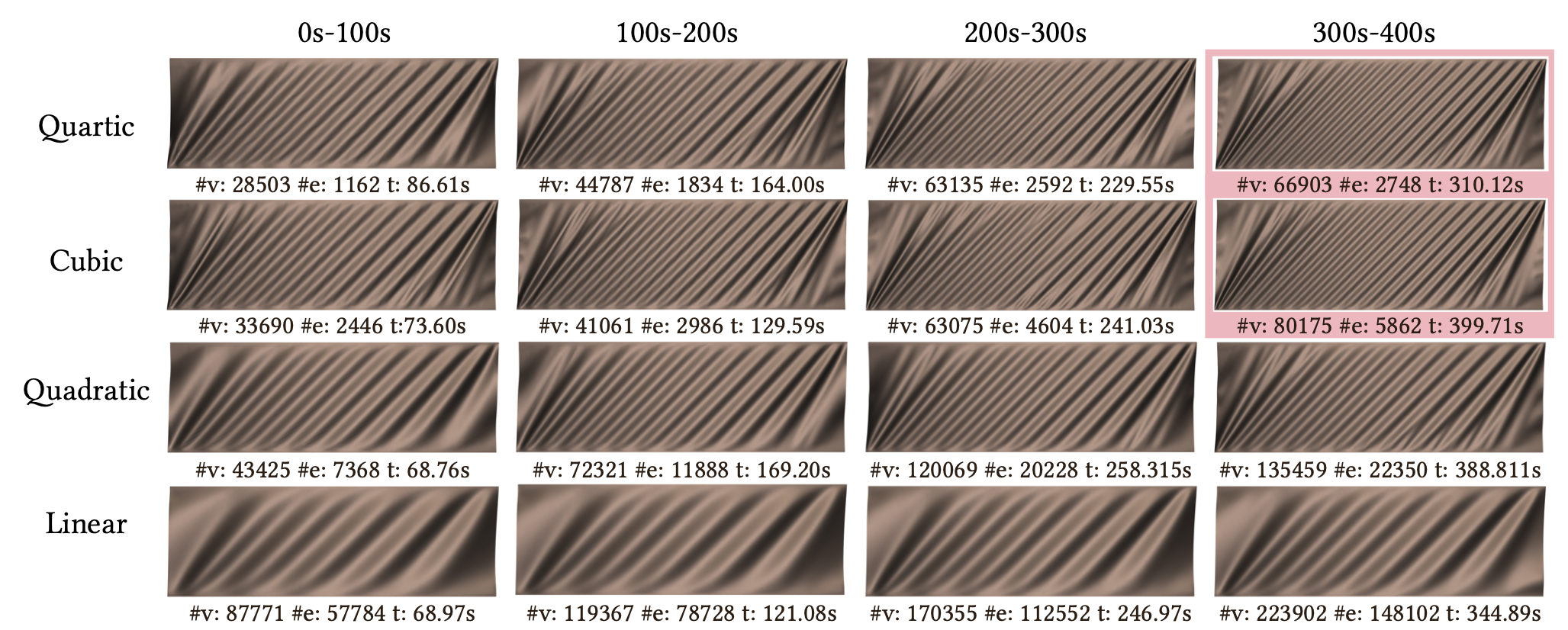}
   \caption{Evolution of wrinkle frequency across interpolation orders. Comparison of wrinkle formation over time (0s to 400s) using linear, quadratic, cubic, and quartic interpolation. While the linear and lower-order formulations predict fewer wrinkles within the given simulation window, the cubic and quartic elements converge toward a consistent high-frequency morphology (highlighted in red). Notably, the quartic formulation reaches this stable wrinkle state slightly faster than the cubic one, demonstrating superior efficiency in capturing complex buckling instabilities within the same time budget.}
   \label{fig:wrinkled-sheet}
\end{figure*}

\rev{To create realistic facial wrinkling, it is essential for our model to capture high frequency wrinkling patterns. To validate our choice of high-order solid shell models for producing high quality wrinkles, we replicate the shearing experiment of Wong and Pellegrino \shortcite{Wong2006}. Our experiment consists of a thin rectangular membrane ($380 \times 128 \times 0.025$ mm) with a Young’s Modulus of $3.5$ GPa and a Poisson’s ratio of $0.31$, and the membrane is subjected to a $3$ mm horizontal shear. This setup is a well-established benchmark for triggering buckling instabilities that should result in a pattern of 19 wrinkles.}

\rev{In \autoref{fig:wrinkled-sheet}, we compare wrinkle frequency across linear, quadratic, cubic and quartic interpolation orders and demonstrate that the higher-order models converge to the expected high-frequency morphology. We evaluate convergence with respect to simulation time
and adjust mesh resolution such that each simulation finishes within the given time window ($0-100$s, $100-200$s, $200-300$s, $300-400$s). While simulations with linear and quadratic elements do not even reach the full wrinkle frequency within the $300-400$s window, cubic and quartic shells converge to the expected high-frequency morphology. Notably, quartic elements reach this stable state slightly faster than their cubic counterparts. This experiment demonstrates that high-order functions combined with L-BFGS offer a promising solution for simulating the smooth, high-frequency displacement distributions necessary to recreate skin wrinkles.}


\subsection{Performance and Statistics}

All simulations were conducted on a MacBook Pro M1 Pro with 16 GB of RAM. \autoref{tab:optPerformance} provides a detailed breakdown of simulation and performance numbers for our experiments. Our experiments demonstrate that L-BFGS consistently outperforms Newton’s method in terms of total computation time. Although L-BFGS typically requires a higher number of iterations to converge, it also avoids the costly, repeated factorization of the system Hessian. For the elastic sheet benchmarks, L-BFGS remains highly competitive and stable even as the interpolation order increases, maintaining consistent timings for similar node counts. In contrast, Newton’s method shows significant performance degradation with higher-order elements. This is due to the increasingly dense Hessian structures associated with quartic and cubic elements, which significantly raise the computational cost of each Newton step.

\begin{table*}[t]
	\centering
	\caption{Summary of experimental setup and performance. Elastic sheet timings reflect the full simulation; skin simulation timings are reported per frame.}
        \begin{tabular}
        {p{2.4cm}p{1.1cm}p{2.0cm}p{2.2cm}p{2.2cm}p{2.1cm}p{1.6cm}}
		\toprule
        Example                      & \#nodes              & \#iter (L-BFGS)      & \# iter (Newton)     & Time (L-BFGS)      & Time (Newton)      & Figure\\
                                     &			           &					  &					     & Per iter / Total	  &	Per iter / Total   &  \\
		\midrule
        Sheet (Linear)      	     & 19620		        & 400 				   & 83				      & 0.029s / 11.17s	   & 0.42s  / 34.98s    & \autoref{fig:wrinkled-sheet} \\
        Sheet (Quadratic)      	     & 20007		        & 1178 				   & 274			      & 0.029s / 34.13s    & 0.56s  / 153.61s   & \autoref{fig:wrinkled-sheet} \\
        Sheet (Cubic)      	         & 18039		        & 1300 				   & 290			      & 0.027s / 34.10s    & 0.62s  / 180.73s   & \autoref{fig:wrinkled-sheet} \\
        Sheet (Quartic)      	     & 18195		        & 1327 				   & 210		          & 0.032s / 42.85s    & 0.83s  / 173.53s   & \autoref{fig:wrinkled-sheet} \\
        Skin (Sphere)      	         & 118902		        & 51 				   & 17		              & 1.27s  / 64.94s	   & 23.10s / 392.66s   & \autoref{fig:ligament-stiffness},\ref{fig:ligament-ratio} \\
        Skin (Temple)      	         & 75159		        & 50 				   & 25		              & 0.85s  / 42.61s	   & 7.44s  / 186.05s   & \autoref{fig:teaser} \\
        Skin (Forehead)      	     & 87879		        & 67 				   & 25		              & 0.96s  / 64.52s	   & 7.08s  / 177.24s   & \autoref{fig:teaser},\ref{fig:temporal} \\
		\bottomrule
	\end{tabular}
	\label{tab:optPerformance}
\end{table*}
\section{Conclusions}


\rev{We presented a facial skin model capable of capturing high-frequency wrinkling details under contact. Our approach relies on three key components: a high-order geometric formulation, a heterogeneous viscoelastic material model, and an anatomically-informed constraint system. Our framework employs high-order prismatic solid-shell elements with quartic in-plane interpolation to resolve bending modes responsible for wrinkles, and quadratic through-the-thickness interpolation, to accurately capture the transverse normal and shear stresses that govern contact kinematics and the non-linear stress gradients induced by finger interactions. We model the multi-layered nature of skin with heterogeneous viscoelastic material parameters at quadrature points with time-dependent relaxation for realistic damped wrinkle formation. Finally, we handle skin attachments to the underlying skeletal structure using spatially varying ligament-like structures that effectively constrain skin mobility. 
By integrating these geometric and anatomical elements into a robust framework, we obtain simulation results that qualitatively match the response of human skin. Our performance analysis demonstrates that combining high-order solid shell elements with an L-BFGS quasi-Newton solver enables superior efficiency and accuracy compared to conventional low-order elements and standard Newton solvers.}


\rev{Although our model effectively captures intricate skin dynamics, several limitations remain. Modeling high-frequency wrinkling comes at significant computational cost, limiting our experiments to localized regions of interest rather than full-head simulations. To mitigate this overhead, we model the multi-layered skin anatomy using a single homogenized solid-shell layer. While effective, this approximation does not explicitly account for inter-layer sliding or the distinct mechanical gradients between the epidermis and the hypodermis. Finally, a significant challenge is the lack of identity-specific anatomical data. Although our results are validated against real-world footage, the spatial placement and density of ligaments were empirically determined. Identity-specific ligament and material parameters could be estimated from capture data through inverse simulation, but our method does not currently support this process.}

\rev{Ultimately, our work creates several exciting directions for future research, specifically in the development of multi-layered shell formulations, the integration of non-invasive imaging data to automate the mapping of ligamentous anchors for personalized facial simulations, and the generation of synthetic data for training digital twins.}

\bibliographystyle{eg-alpha-doi}  
\bibliography{reference.bib}

\end{document}